\documentclass{aa}
\usepackage{graphics}

\newcommand{\Dc}{\Delta_{\rm c}}
\newcommand{\omm}{\Omega_{\rm M}}

\begin{document}
 
\thesaurus{02 (3.13.18) - methods: N-body simulations}
\title{The mass of a halo}
\author{Martin White}
\institute{Harvard-Smithsonian Center for Astrophysics, Cambridge, MA 02138}
\date{Received: XXX; accepted XXX}

\maketitle

\begin{abstract}
We discuss the different definitions of the mass of a halo in common
use and how one may convert between them.  Using N-body simulations we
show that mass estimates based on spherical averages are much more tightly
correlated with each other than with masses based on the number of particles
in a halo.
The mass functions pertaining to some different mass definitions are estimated
and compared to the `universal form' of Jenkins et al.~(\cite{JFWCCEY}).
Using a different simulation pipeline and a different cosmological model we
show that the mass function is well fit by the Jenkins et al.~(\cite{JFWCCEY})
fitting function, strengthening the claim to universality made by those
authors.
We show that care must be taken to match the definitions of mass when
using large N-body simulations to bootstrap scaling relations from smaller
hydrodynamical runs to avoid observationally significant bias in the
predictions for abundances of objects.
\end{abstract}

\section{Introduction}

One of the most fundamental predictions of a theory of structure formation
is the number density of objects of a given mass, the mass function, at a
given redshift.  Accurate mass functions are used in a number of areas in
cosmology; in studies of galaxy formation, in measures of volumes (e.g.~galaxy
lensing) and in attempts to infer the normalization of the power spectrum and
the density parameter from the abundance of rich clusters.
In the latter case the mass function is the point of contact allowing us to
bootstrap the excellent statistics of large N-body simulations with observable
properties of clusters normalized for example by hydrodynamic simulations of
smaller volumes.  In this way we can obtain reliable estimates of
e.g.~the number of clusters as a function of temperature and redshift, which
can in turn be used to constrain the matter density, the normalization of the
power spectrum and the statistics of the initial density field.

Several different definitions for the ``mass'' of a halo are in common use,
each having different advantages.
For example there are at least 3 different definitions of mass for the
mass-temperature relation as computed from hydrodynamical simulations of
galaxy clusters and all are different from the definition of mass commonly
used in the mass function which is itself different from the mass usually
employed in analytic studies based on the Press-Schechter~(\cite{PS};
hereafter PS) theory.
Observational data have improved to the point where it is important to
distinguish between these different definitions of mass, lest we bias
our theoretical predictions.  We give an example of how such a
conversion can be made, at least approximately, for a certain class of mass
estimator.

\section{Halo mass definitions}

In this section we describe some of the definitions of the mass of a dark
matter halo in common use.  This list is obviously not exhaustive, but it is
representative.
We begin by recalling some background about the spherical top-hat collapse
model (see e.g.~Peebles~\cite{PPC}; Peacock~\cite{Peacock};
Liddle \& Lyth~\cite{LidLyt}, and references therein) from which much of
the language in this field has been borrowed.

The spherical top-hat ansatz describes the formation of a collapsed object
by solving for the evolution of a sphere of uniform overdensity $\delta$ in
a smooth background of density $\bar{\rho}$.  By Birkhoff's theorem the
overdense region evolves as a positively curved Friedman universe whose
expansion rate is initially matched to that of the background.
The overdensity at first expands but because it is overdense the expansion
slows (relative to the background) and eventually halts before the region
begins to recollapse.
Technically the collapse proceeds to a singularity but it is assumed in a
``real'' object virialization occurs at twice\footnote{There is a small
correction to this in the presence of a cosmological constant which
contributes a $\Lambda r^2$ potential.}
the turn-around time, resulting in a sphere of half the turn-around radius.
In an Einstein-de Sitter model the overdensity (relative to the critical
density) at virialization is $\Delta_{\rm c}=18\pi^2\simeq 178$.  We shall
always use $\Delta_{\rm c}$ to indicate the overdensity relative to critical
of a virialized halo, which will be lower for smaller $\Omega_{\rm M}$.
Note that some authors use a different convention in which $\Dc$ is specified
relative to the background matter density -- our $\Dc$ is $\omm$ times theirs.
The linear theory extrapolation of this overdensity is normally denoted
$\delta_{\rm c}$ and is $(3/20)(12\pi)^{2/3}\simeq 1.686$ in an
Einstein-de Sitter model.  This overdensity is often used as a threshold
parameter in PS theory and its extensions and has a very weak cosmology
dependence which is often neglected.

\begin{figure}
\begin{center}
\resizebox{3.5in}{!}{\includegraphics{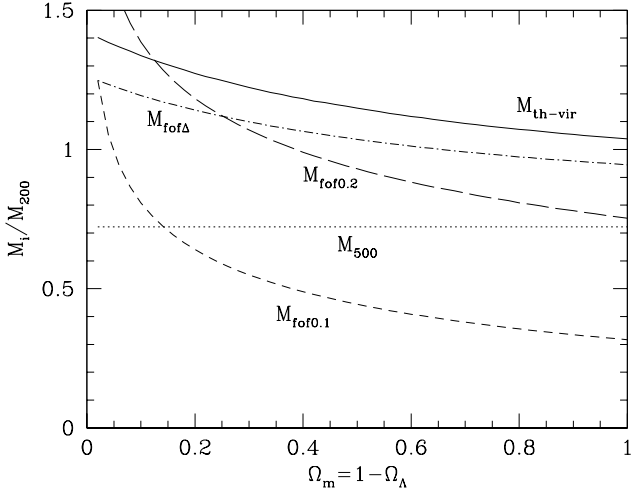}}
\end{center}
\caption{The relations between various definitions of the mass of a halo
as a function of $\Omega_{\rm M}$ assuming the halo density profile follows the
NFW form with concentration parameter $c=5$.}
\label{fig:mdef}
\end{figure}

We now turn to some of the definitions of mass in use in the literature.
Let us suppose that we have identified some object, a problem to which we
shall return in \S\ref{sec:sims}, and have chosen a fiducial `center' about
which to take spherically averaged profiles.  We define $M_\Delta$ as the
mass contained within a radius $r_\Delta$ inside of which the mean interior
density is $\Delta$ times the {\it critical\/} density
\begin{equation}
  \int_0^{r_\Delta} r^2 dr\ \rho(r) =
    {\Delta\over 3} \rho_{\rm crit} r_\Delta^3
  \qquad .
\end{equation}
The `virial mass' from the spherical top-hat collapse model would then be
simply $M_{\Delta_{\rm c}}$.  We shall refer to this mass as
$M_{{\rm th-vir}}$.

An alternative interpretation of the spherical collapse model is that the
virial radius corresponds to that point within which the material is
virialized and external to which the mass is still collapsing onto the
object.  Some simulations suggest that this occurs at $\Delta=200$ more
or less independent of cosmology, and so a common mass estimator is
$M_{200}$ which is approximately the virial mass if $\Omega_{\rm M}=1$.
Authors also use $M_{500}$ or $M_{1000}$ in some applications.

\begin{figure}
\begin{center}
\resizebox{3.5in}{!}{\includegraphics{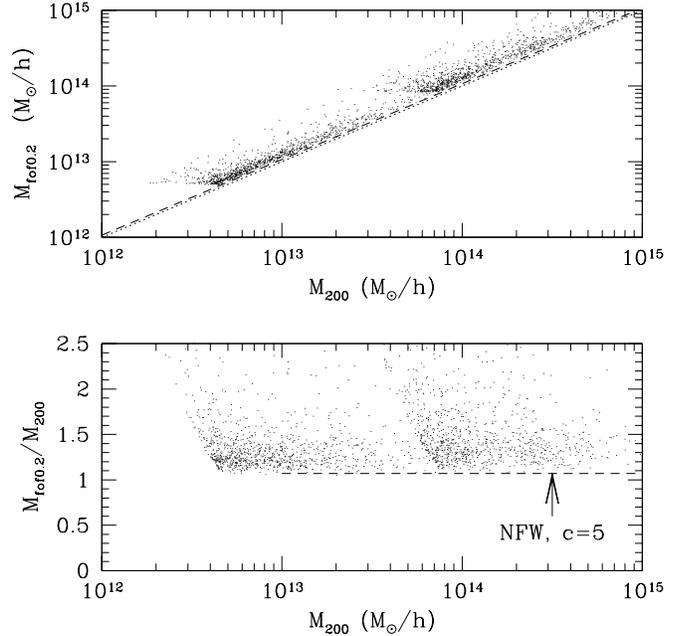}}
\end{center}
\caption{Correlations between two estimators of the mass, $M_{{\rm fof}0.2}$
and $M_{200}$, in simulations of a $\Lambda$CDM model (see text).  Only halos
with more than 1000 particle are used.  The upper panel shows that the
different estimates of mass are correlated, but offset (the dotted line
marks $y=x$).  The dashed line indicates the prediction for the offset of
\S\protect\ref{sec:corr}.  The lower panel shows the mass ratios as a
function of $M_{200}$ on a linear scale.}
\label{fig:mcorr}
\end{figure}

The final mass estimator we shall consider is the one used by
Jenkins et al.~(\cite{JFWCCEY}) in their analysis of simulations performed
by the Virgo consortium.  The mass they assign to a halo is simply the sum
of the masses of the particles identified as members of the halo by their
halo finder.  This mass estimator is the least easily interpreted
theoretically, though it is well defined algorithmically, but has the very
convenient feature that the mass function for this estimator is independent
of cosmology!

Unfortunately there is no unique algorithmic definition of a dark matter halo,
even within a 3D simulation itself.
To define the mass of an object we must first specify the object in question!
Although other halo finders are in common use, we shall deal exclusively with
halos found using the Friends-of-Friends (Davis et al.~\cite{DEFW}) algorithm,
hereafter called FOF.  Our reason for this choice is that it is the group
finder employed by the Virgo consortium in deriving the mass functions from
their very large N-body simulations (Jenkins et al.~\cite{JFWCCEY}) and in
terms of which they find a universal mass function.
The FOF algorithm has one free parameter, $b$, the linking length in units
of the mean inter-particle spacing.
Commonly used values of $b$ are 0.1, 0.15 and 0.2, although other choices
exist (e.g.~Gardner~\cite{Gar} chooses instead
$b^{-3}=\Omega_{\rm M}\Delta_{\rm c}/3$).  Jenkins et al.~(\cite{JFWCCEY})
find that the mass function is universal if they take $b=0.2$, independent
of the cosmology under consideration.
Jenkins et al.~(\cite{JFWCCEY}) claim that in the limit of a large number of
particles per halo, FOF finds all particles within an iso-density contour
$b^{-3}$ times the mean matter density or $\Omega_{\rm M}/b^3$ times critical
density.  We shall return to this point in \S\ref{sec:sims}.

In the next sections we describe how these different masses are related.

\begin{figure*}
\begin{center}
\resizebox{6in}{!}{\includegraphics{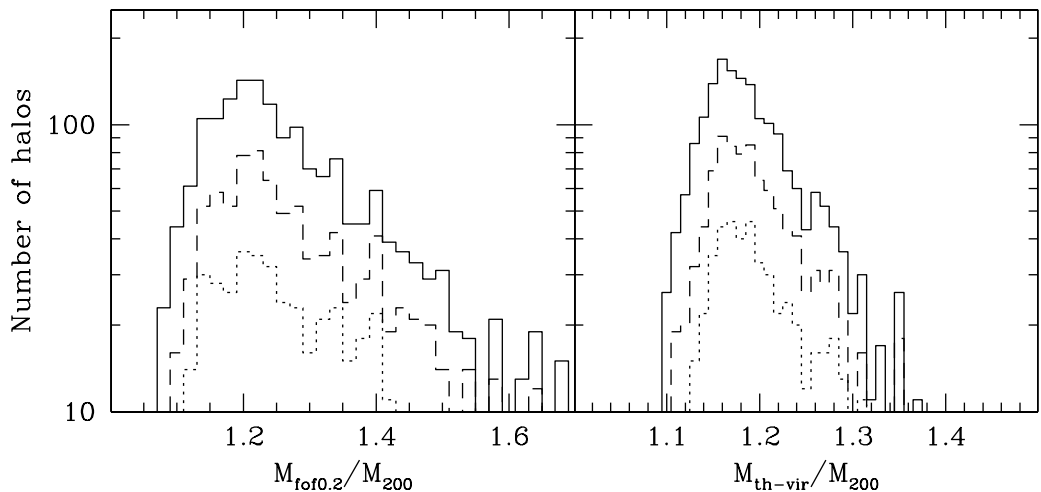}}
\end{center}
\caption{Distribution of mass ratios for halos from the first simulation.
The solid line shows all halos with more than 500 particles, the dashed
line 1000 particles and the dotted line 2000 particles.  The ratios predicted
by the model of \S\protect\ref{sec:corr} are $M_{{\rm fof}0.2}/M_{200}=1.07$
and $M_{\rm th-vir}/M_{200}=1.22$.}
\label{fig:hist}
\end{figure*}

\section{An approximate conversion} \label{sec:corr}

It would be advantageous if we could convert, at least approximately, between
the different definitions of mass in common use.  In this section we make a
first attempt in this direction.  We assume that all halos are spherical and
have a universal profile, for example\footnote{There is some controversy in
the literature about the slope of the density profile as $r\to 0$.  Luckily
departures from the NFW form are expected to occur only at very small radii,
containing a vanishingly small amount of the total mass, and thus do not
concern us here.} the NFW form (Navarro, Frenk \& White~\cite{NFW}):
\begin{equation}
  \rho(r) \propto x^{-1}\left(1+x\right)^{-2},
\end{equation}
where $x=r/r_{\rm s}$ and $r_{\rm s}$ is a scale radius usually specified
in terms of the concentration parameter $c\equiv r_{200}/r_{\rm s}$.
Navarro et al.~(\cite{NFW}) refer to $r_{200}$ and $M_{200}$ throughout as
the `virial radius' and `virial mass' respectively.
Again, N-body simulations have shown that the concentration parameter is
a weak function of virial mass, taking the value $c\sim 5$ for masses
characteristic of clusters which shall be our focus here.

With a given profile it is straightforward to relate the various mass
definitions as shown in Fig.~\ref{fig:mdef}.  Note that the masses can
differ significantly in this model and that the cosmology dependence is
quite strong.  How well does this crude model predict the relationship between
different mass definitions in practice?

\begin{table}
\begin{center}
\begin{tabular}{c|ccc}
                   & $M_{{\rm fof}0.2}$ & $M_{\rm th-vir}$ & $M_{200}$ \\
\hline
$M_{{\rm fof}0.2}$ & 1.000 & 0.974 & 0.966 \\
$M_{{\rm th-vir}}$ & 0.974 & 1.000 & 0.997 \\
$M_{200}$          & 0.966 & 0.997 & 1.000
\end{tabular}
\end{center}
\caption{Correlations between the various mass definitions for the halos
of our large simulation.  Only halos with more than 1000 particles are used.}
\label{tab:mcov}
\end{table}

\section{Simulations} \label{sec:sims}

The model of the previous section made several approximations.
To compare these different mass estimates with each other more carefully we
have used some N-body simulations, originally run for another purpose.
We simulated the Ostriker \& Steinhardt~(\cite{OstSte}) concordance model,
which has $\Omega_{\rm m}=0.3$, $\Omega_\Lambda=0.7$,
$H_0=100\,h\,{\rm km}{\rm s}^{-1}{\rm Mpc}^{-1}$ with $h=0.67$,
$\Omega_{\rm B}=0.04$, $n=1$ and $\sigma_8=0.9$
(corresponding to $\delta_H=5.02\times 10^{-5}$).
For this cosmology $\Delta_{\rm c}\simeq 101$ from the top-hat collapse model.

We have used two `high' resolution and two lower resolution N-body
simulations.
The first was a $256^3$ particle, dark matter only simulation in a
$100^3\,h^{-1}$Mpc box.
The simulation was run on 16 processors of the Origin2000 at NCSA
with the {\sl TreePM-SPH\/} code (White et al.~\cite{TreePM}) operating
in collisionless (dark matter only) mode.
The gravitational force softening was of a spline form
(e.g.~Hernquist \& Katz~\cite{HerKat}), with a ``Plummer-equivalent''
softening length of $15\,h^{-1}{\rm kpc}$ comoving.
The simulation was evolved from $z=100$ until the present and took a
total of 1300 (normalized) CPU hours.
A second $256^3$ particle simulation was run from $z=70$ in a $256\,h^{-1}$Mpc
box with a $35\,h^{-1}$kpc softening.  This run took a total of 880
(normalized) CPU hours, also on 16 processors.
To provide additional statistics on the high-mass end of the mass function we
additionally ran several smaller ($128^3$ particles) simulations of the same
model in boxes of side $200\,h^{-1}$Mpc.

{}From the $z=0$ outputs we generated group catalogues using the FOF
algorithm with $b=0.2$.  For each halo we defined the center for our
spherical averages as the particle with the minimum potential energy.
This corresponds closely to the most bound particle and the density
peak for a halo in all but the most disturbed systems, and is more
robust than the center of mass.  With the group catalog and centers
so defined it is straightforward to calculate each of the estimators
described above.
Throughout we shall use $M_{200}$ as our ``base'' mass to which the others
are compared, since this fits best into the philosophy of \S\ref{sec:corr}.

We show in Fig.~\ref{fig:mcorr} that the different mass estimators are indeed
highly correlated as one would expect (see also Table~\ref{tab:mcov}).
The ratio of the FOF mass to $M_{200}$ is seen to span a reasonable range and
be offset from unity by a non-negligible amount.
Fig.~\ref{fig:hist} shows a histogram of the ratio of $M_{{\rm fof}0.2}$
and $M_{\rm th-vir}$ to $M_{200}$ for all of the halos above a given number
of particles.

Let us focus first on $M_{\rm th-vir}$.
The shape of the histogram is roughly the same regardless of particle number
indicating that the distribution is not strongly affected by finite particle
numbers in the halos chosen.  The top-hat virial mass and $M_{200}$ are fairly
tightly correlated, as one might expect since they are both spherically
averaged statistics and the density contrasts are not too different.
The approximation of \S\ref{sec:corr} provides a reasonable estimate of
the ratio of masses, presumably because of the spherical averaging being
performed.

\begin{figure}
\begin{center}
\resizebox{3.5in}{!}{\includegraphics{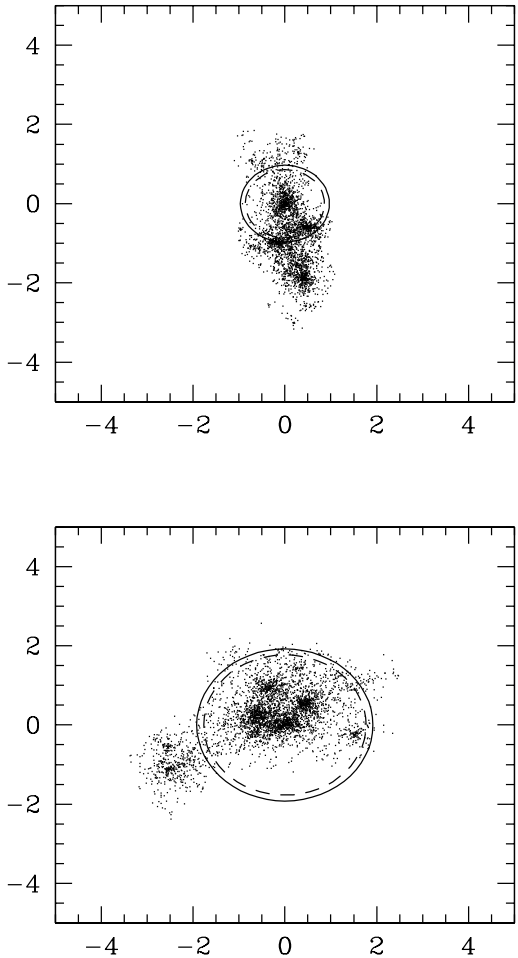}}
\end{center}
\caption{The mass distribution for two groups from the first simulation,
along with the top-hat `virial' radius (solid) and $r_{200}$ (dashed line).
An $x-y$ projection, with axes in $h^{-1}$Mpc, is shown centered on the
point of minimum potential.
The upper panel shows a group with $M_{200}\simeq 10^{14}h^{-1}M_\odot$
(62,000 particles; 5\% shown), while the lower panel shows the largest
group in the simulation ($M_{200}=8\times 10^{14}h^{-1}M_\odot$, $2\times10^5$
particles; 2\% shown).
We have plotted only a fraction of the points for clarity, this suppresses
the appearance of substructure in the group.
There is a tendency for the larger groups to be more irregular.}
\label{fig:dotplot}
\end{figure}

The situation with regards $M_{{\rm fof}0.2}$ is more complicated.  Firstly
the scatter is much larger, since the groups found by FOF can be quite
irregularly shaped (see Fig.~\ref{fig:dotplot}).
The actual ratio of $M_{{\rm fof}0.2}$ to $M_{200}$ is larger than predicted
by our model.  By eye the halo finder looks to have merged close groups.
It is possible that with better mass resolution this would occur to a lesser
extent, although the lack of a clear trend with particle number in
Fig.~\ref{fig:hist} argues that convergence would be quite slow.

To investigate whether FOF has merged adjacent halos and whether a FOF halo
can be considered as all particles above a given density threshold (as
assumed in \S\ref{sec:corr}) we have examined the particles in the two groups
of Fig.~\ref{fig:dotplot}.
For each particle we calculate its distance from the halo center and its
density, defined from the distance ($r_{\rm neig}$) to the 32nd nearest
neighbour\footnote{We use all of the particles in the simulation, not just
group members, in computing $r_{\rm neig}$.} as
\begin{equation}
  \rho \equiv {3M\over 4\pi} r_{\rm neigh}^{-3}
\label{eqn:dens32}
\end{equation}
where $M$ is the enclosed mass.  This density estimate is somewhat noisy, but
will be sufficiently accurate for our purposes.

\begin{figure}
\begin{center}
\resizebox{3.5in}{!}{\includegraphics{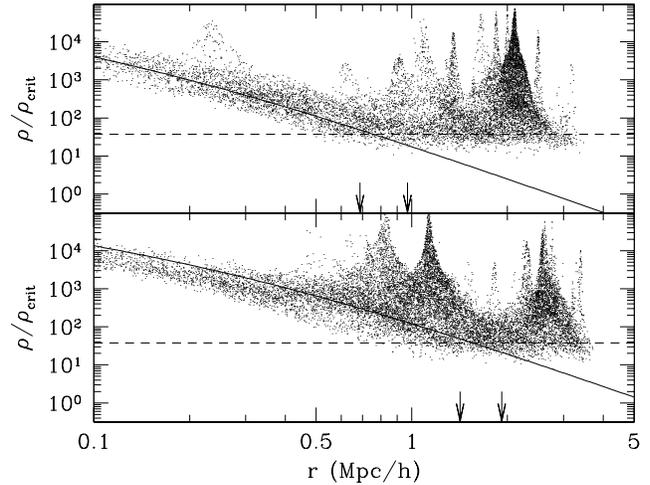}}
\end{center}
\caption{The halo density profile, as a function of radial distance from the
group center, for the groups of Fig.~\protect\ref{fig:dotplot}.  Each dot
represent a particle in the group (only a small fraction of the particles are
shown for clarity) and the density is estimated through
Eq.~\protect\ref{eqn:dens32}.  The dashed horizontal line is $\omm b^{-3}$,
the arrows mark $r_{200}$ and $r_{\rm th-vir}$ and the solid line is the
`model' NFW profile of \S\protect\ref{sec:corr}, with $c=5$.}
\label{fig:rdens}
\end{figure}

The results are shown in Fig.~\ref{fig:rdens} along with the appropriate
NFW profile (assuming $c=5$) for comparison.  Note that the substructure
in the halos is readily apparent as density `spikes' as functions of
radius and can be matched to `sub-halos' in Fig.~\ref{fig:dotplot}.

It is clear from Fig.~\ref{fig:rdens} that FOF is (at least approximately)
linking all particles above a density $\Omega_{\rm M}b^{-3}$ of critical.
However the chosen contrast is low enough that this includes particles well
outside $r_{200}$ (or the virial radius) and the presence of these `nearby'
particles is biasing the masses above $M_{200}$ and contributing to the
extra scatter.  This problem could be mitigated to some extent by reducing
$b$, but it is not clear whether the mass function would remain universal if
a different (cosmology independent) $b$ were chosen.

\begin{figure}
\begin{center}
\resizebox{3.5in}{!}{\includegraphics{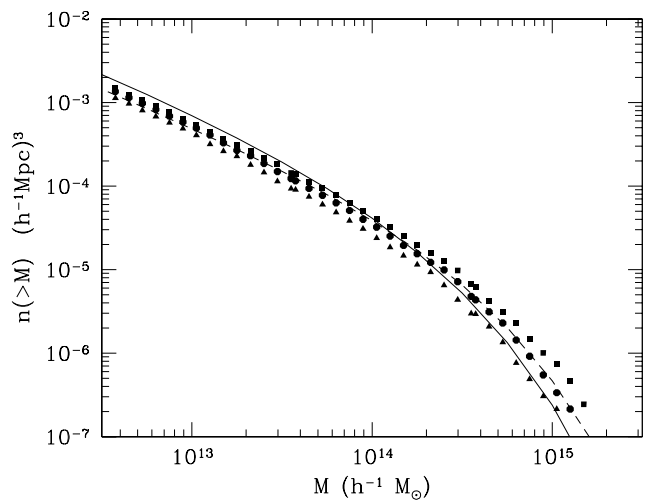}}
\end{center}
\caption{The cumulative mass function for the Concordance model.  The points
are mass functions from the simulations where the mass is defined as:
FOF0.2 (squares), $M_{\rm th-vir}$ (circles) and $M_{200}$ (triangles).
The solid line shows the predictions of the Press-Schechter theory, the
dashed line is the fit of Jenkins et al.~(\protect\cite{JFWCCEY}).}
\label{fig:massfn}
\end{figure}

Finally we show in Fig.~\ref{fig:massfn} the different mass functions that
would be derived from these halos using the 3 different mass definitions.
Note that our results, obtained with a completely independent N-body code
and analyzed with a completely independent set of software, agree well
with the universal mass function of Jenkins et al.~(\protect\cite{JFWCCEY}).
However, especially at the high mass end, the number density is quite
sensitive to the mass definition used.  Intriguingly the fitting formula
quoted by Jenkins et al.~(\protect\cite{JFWCCEY}) provides a good fit to
the mass function if we use $M_{{\rm th-vir}}$ as our definition of mass,
an even better fit than if we use $M_{{\rm fof}0.2}$.

\section{An example: the temperature function}

An example of where this mass uncertainty may be important is in using
the observed temperature function of local X-ray clusters to constrain
the normalization of the present day power spectrum.
Although Pen~(\cite{Pen}) has argued for a direct prediction of the
temperature function from simulations, most people use a hybrid approach
in obtaining this constraint.  First a mass function, usually calibrated by
simulations such as those of Jenkins et al.~(\cite{JFWCCEY}), is assumed
and then a mass temperature relation from hydrodynamic simulations is
used to predict the temperature function.

Hydrodynamic simulations show good agreement for the total mass and X-ray
temperature properties of clusters (e.g.~Frenk et al.~\cite{Frenketal}).
The mass-temperature relation from these simulations generally follows
the virial relation quite well.
Ignoring a small correction from the $\Lambda r^2$ potential,
for an object virialized at a redshift $z$
\begin{equation}
  M(T,z) \propto \left[ T/(1+z) \right]^{3/2}
\label{eqn:m-t}
\end{equation}
and the proportionality constant is then fixed by the simulations.
There still remains some disagreement over the precise coefficient and its
dependence on cosmology (see e.g.~Henry~(\cite{Henry}) for a recent
compilation) but in addition to this different authors choose different
conventions for $M$.  For reasonable $\omm$, the difference in mass
definitions is about the same size as the difference in proportionality
constants between groups and the $1\sigma$ scatter in $M$ at fixed $T$ found
by the different groups.
Unfortunately this alone does not account for all of the discrepancy.

To take only two relations as examples, Evrard, Metzler \& Navarro~(\cite{EMN})
define the mass in Eq.~\ref{eqn:m-t} as $M_{200}$.
The $M$ of Bryan \& Norman~(\cite{BryNor}), on the other hand, is the
`virial' mass in the sense of the spherical top-hat model.
In both cases there is ample justification for the definition chosen, and
the authors are explicit about their choice.  However, there is an $\omm$
dependent ratio between these two definitions which cannot be neglected if
the temperature function is to be predicted to the factor of 2 level!

\section{Conclusions}

We have shown that different definitions of the ``mass'' of a halo exist,
and have different strengths and weaknesses.  It is important to be consistent
when combining relations which use different definitions of mass, and we
have given an approximate method for converting between some commonly used
mass estimators.
Mass estimates based on spherical averages are much more tightly correlated
with each other than with the mass obtained simply by summing the particles
in the group, and can be quite well estimated by assuming a `universal'
spherical profile (\S\ref{sec:corr}).
The hydrodynamical simulations which calibrate observables as a function of
cluster mass typically use such a spherically averaged mass definition.

Unfortunately these definitions are not very tightly correlated with the
particle based mass used in the universal mass function of halos reported
by Jenkins et al.~(\cite{JFWCCEY}).  We have argued that this is because,
with a linking length of $b=0.2$, FOF is merging neighbouring halos.  Such
a problem would be mitigated by reducing $b$, but it has not been demonstrated
that the mass function is universal for $b\ne 0.2$.

In the cosmological model we have simulated, the fitting form of
Jenkins et al.~(\cite{JFWCCEY}) provides a good match to the mass function of
our halos, strengthening the claim of those authors that it is universal.
However we find the best fit when the mass estimator used is the top-hat
virial mass, rather than the FOF mass.
While we have not investigated other cosmological models, we expect that the
profiles (from the point of view of estimating masses) will not be very
cosmology dependent.
It would be interesting to see whether the Jenkins et al.~(\cite{JFWCCEY})
mass function is `universal' if one uses the (cosmology dependent) top-hat
virial mass of the halos.

\bigskip
I would like to thank V.~Springel for the use of his FOF group finder,
J.~Mohr for useful conversations on this issue, and C.~Metzler for
encouraging me to write it up.
M.~White was supported by the US National Science Foundation and a
Sloan Fellowship.
Parts of this work were done on the Origin2000 system at the National
Center for Supercomputing Applications, University of Illinois,
Urbana-Champaign.

\end{document}